\documentstyle[titlepage,12pt]{article}
\begin{document}
\newfont{\bg}{cmr10 scaled\magstep3}
%
\newcommand{\gsimeq}
{\hbox{ \raise3pt\hbox to 0pt{$>$}\raise-3pt\hbox{$\sim$} }}
\newcommand{\lsimeq}
{\hbox{ \raise3pt\hbox to 0pt{$<$}\raise-3pt\hbox{$\sim$} }}
\newcommand{\plb}[3]{Phys. Lett. {\bf B#1}, #3 (#2)}
\newcommand{\prl}[3]{Phys. Rev. Lett. {\bf #1}, #3 (#2)}
\newcommand{\prd}[3]{Phys. Rev. {\bf D#1}, #3 (#2)}
\newcommand{\npb}[3]{Nucl. Phys. {\bf B#1}, #3 (#2)}
\newcommand{\prog}[3]{Prog. Theor. Phys. {\bf #1}, #3 (#2)}
\newcommand{\zeitc}[3]{Z. Phys. {\bf C#1}, #3 (#2)}
\newcommand{\mpl}[3]{Modern. Phys. Lett. {\bf A#1}, #3 (#2)}
\newcommand{\ijmp}[3]{Int. J. Modern. Phys. Lett. {\bf A#1}, #3 (#2)}
\newcommand{\bbar}{\bar{B}}
\newcommand{\kstar}{{K^{\ast}}}
\newcommand{\dstar}{{D^{*}}}
\newcommand{\lbar}{\bar{l}}
\newcommand{\mw}{m_{W}}
\newcommand{\mt}{m_{t}}
\newcommand{\mb}{m_{b}}
\newcommand{\mcc}{m_{c}}
\newcommand{\ms}{m_{s}}
\newcommand{\mbbar}{m_{\bbar}}
\newcommand{\mrho}{m_{\rho}}
\newcommand{\mkstar}{m_{\kstar}}
\newcommand{\bsll}{b \rightarrow s l \lbar}
\newcommand{\bksll}{\bbar \rightarrow \kstar l \lbar}
\newcommand{\brln}{\bbar \rightarrow \rho l \bar{\nu}}
\newcommand{\bdln}{\bbar \rightarrow \dstar l \bar{\nu}}
\newcommand{\bksg}{\bbar \rightarrow \kstar \gamma}
\newcommand{\vub}{V_{ub}}
\newcommand{\vtb}{V_{tb}}
\newcommand{\vsts}{V^{\ast}_{ts}}
\newcommand{\sbar}{\bar{s}}
\newcommand{\gf}{G_{F}}
\newcommand{\rttwo}{\sqrt{2}}
\newcommand{\epstar}{{\epsilon^{\ast}}}
\newcommand{\epp}{\epstar\cdot p}
\newcommand{\pppp}{p+p'}
\newcommand{\pmpp}{p-p'}
\newcommand{\qsq}{q^2}
\newcommand{\qsqmax}{{q^2_{max}}}
\newcommand{\qsqmr}{{\qsq}^{\bbar \rightarrow \rho}_{max}}
\newcommand{\qsqmk}{{\qsq}^{\bbar \rightarrow \kstar}_{max}}
\newcommand{\prho}{\mbox{\boldmath $p$}_{\rho}}
\newcommand{\vprho}{\vec{\mbox{\boldmath $p$}}_{\rho}}
\newcommand{\pks}{\mbox{\boldmath $p$}_{\kstar}}
\newcommand{\vpks}{\vec{\mbox{\boldmath $p$}}_{\kstar}}
\newcommand{\ap}{a_{+}}
\newcommand{\am}{a_{-}}
\newcommand{\gmu}{\gamma_{\mu}}
\newcommand{\gnu}{\gamma_{\nu}}
\newcommand{\gfive}{\gamma_{5}}
\newcommand{\hmrb}[1]{<\rho(p',\epsilon)|\bar{u}{#1}b|\bbar(p)>}
\newcommand{\hmkb}[1]{<\kstar(p',\epsilon)|\bar{s}{#1}b|\bbar(p)>}
\begin{titlepage}
\rightline{DPNU-95-17}
\rightline{UT-709}
\rightline{June 1995}
\vspace*{2cm}
\addtocounter{footnote}{1}
\begin{center}
{\large \bf
A new method of determining $|\vub|$ \\
\vspace{0.5cm}
by the processes $\brln$ and $\bksll$
}
\\
\vspace{0.5cm}
\vspace{1cm}
\vspace{2cm}
{\sc A.~I.~Sanda}
\\
\vspace{1cm}
{\it Department of Physics,  Nagoya University \ \
\\
       Chikusa-ku, Nagoya, 464-01 Japan \\}
\vspace{0.5cm}
and \\
\vspace{0.5cm}
{\sc Atsushi Yamada}
\\
\vspace{1cm}
{\it Department of Physics,  University of Tokyo \ \
\\
       Bunkyo-ku, Tokyo, 113 Japan \\}
\vspace{2cm}
{\bf ABSTRACT}
\end{center}
The differential decay width
of the process $\brln$ is related to that of the process
$\bksll$ by using  $SU(3)$-flavor symmetry and the heavy
quark symmetry.
The ratio of the Kobayashi-Maskawa matrix elements is
obtained
in the
zero recoil limit of $\rho$ and $\kstar$,
allowing a determination of $|\vub|$.
\end{titlepage}
\baselineskip = 0.7cm
A precise test of unitarity of the Kobayashi-Maskawa matrix \cite{km}
is essential for further investigations of the quark mass
matrix  and understanding the origin of CP violations.
This is most conveniently performed in the B-meson
systems because of the large CP-violation predicted in this system
\cite{sanda1}.
Strategies for accurate determination
of  Kobayashi-Maskawa matrix elements in B decay is required.
An important number is one of the sides of the unitarity triangle $\vub$
\cite{vubc}.
A well known method is to study the leptonic spectrum at the kinematical
point where the charm quark can not be produced. As there are always
questions as to what extent the obtained result is independent of theoretical
interpretations, it is important to get at the number in as many independent
ways as possible.
In this Letter, we propose a strategy to get at $\vub$.
We shall compliment the analysis of the leptonic spectrum, and propose
to obtain $\vub$ from $\brln$ and $\bksll$. Our analysis uses
$SU(3)$-flavor and heavy quark symmetries.

Consider the zero recoil limit of the $\kstar$ and $\rho$ mesons.
Using heavy quark symmetry,
it will be shown that
the matrix element of the hadronic currents describing
the decay $\bksll$ can be  expressed by
the same  form factor appearing in
the decay $\brln$.
The form factors in each process are
equated the use of $SU(3)$-flavor symmetry, and
the ratio of the Kobayashi-Maskawa matrix elements is
obtained by the ratio of these differential decay widths.
First consider the semileptonic decay $\brln$.
This process is described by the invariant amplitude
\begin{eqnarray}
M=\frac{4\gf}{\rttwo}\vub \bar{u}_L \gmu b_L \lbar_L \gamma^\mu \nu.
\label{eqn:mr}
\end{eqnarray}
The hadronic matrix elements required for this process are
\begin{eqnarray}
\hmrb{\gamma_{\mu}} &=& i g^{\rho} \epsilon_{\mu\nu\rho\sigma}
\epstar^{\nu}(\pppp)^{\rho}(\pmpp)^{\sigma},
\label{eqn:f2}
\\
\hmrb{\gamma_{\mu}\gfive} &=& f^{\rho} \epstar_{\mu}
+\ap^{\rho}(\epp)(\pppp)_{\mu}
+\am^{\rho}(\epp)(\pmpp)_{\mu}.
\label{eqn:f1}
\end{eqnarray}
The form factors $f^\rho$, $a^\rho_{\pm}$ and $g^\rho$
are Lorentz invariant functions of the invariant mass squared
$\qsq=(\pmpp)^2$ of the two leptons.
The $\rho$ meson polarization vector $\epsilon^*$ is given by
\begin{eqnarray}
\epsilon_L&=&(\frac{\prho}{m_\rho},0,0,\frac{E_\rho}{m_\rho}),
\hspace{0.5cm}
\epsilon_\perp=(0,\epsilon_x,\epsilon_y,0).
\end{eqnarray}
Hereafter we study the decay in the rest frame of the $\bbar$-meson, so that
$p^\mu=(\mbbar,\vec{{\bf 0}})$,
$p'^\mu=(E_\rho, \vprho)$ and
the momentum of
the $\rho$ meson $\prho=|\vprho|$ is given by
\begin{eqnarray}
{\prho}=\frac{1}{2\mbbar}
[(\mbbar^2-\mrho^2-\qsq)^2-4\mrho^2\qsq]^{\frac{1}{2}}.
\label{eqn:pr}
\end{eqnarray}
In this frame,
either of
$(\pppp)^\rho$ and $(\pmpp)^\sigma$ in eq. (\ref{eqn:f2}) should be the
spatial components $\pm{\vprho}^i$, so
the right-hand side of eq.(\ref{eqn:f2}) is proportional to $\prho$.
Furthermore, since
$\epp=\epsilon^{\ast0}\mbbar$
and only the longitudinal polarization vector $\epsilon^L$ has a non-zero
time component $\prho/\mrho$, the second and third terms in eq.
(\ref{eqn:f1}) are also proportional to $\prho$.
Thus, the hadronic matrix elements are expanded as
\begin{eqnarray}
\hmrb{\gamma_{\mu}} &=& {\cal O}(\prho),
\label{eqn:fl2}
\\
\hmrb{\gamma_{\mu}\gfive} &=& f^\rho \epstar_{i} + {\cal O}(\prho),
\label{eqn:fl1}
\end{eqnarray}
in the vicinity of the zero recoil $\rho$ meson; $\prho \simeq 0$
, or equivalently, the maximum $\qsq$;
$\qsq \simeq \qsqmax=(\mbbar-\mrho)^2$.
The $\qsq$-distribution of the decay width is computed as
\begin{eqnarray}
\frac{d \Gamma(\brln)}{d\qsq}=
|\vub|^2 {\frac{\gf}{32\pi^3\mbbar^2}}|f^\rho|^2 \qsq \prho
+{\cal O}(\prho^3).
\label{eqn:qsq1}
\end{eqnarray}
Next we consider the flavor changing neutral decay $\bksll$.
In the standard model, this decay
takes place at the loop level via
penguin and box diagrams \cite{inamilim}. The QCD corrected
effective Hamiltonian describing this decay is
\begin{eqnarray}
H&=& \frac{4\gf}{\rttwo\pi}\vtb\vsts
\{
C_7(\mb)O_7 + C^{eff}_8(\mb)O_8 + C_9(\mb)O_9
\}, \label{eqn:hk1}
\end{eqnarray}
where the operators $O_7$, $O_8$ and $O_9$ are defined as
\begin{eqnarray}
O_7=\frac{e^2}{16\pi^2}\mb
\sbar_L i\sigma_{\mu\nu}(q^\nu/\qsq)b_R \lbar \gamma^\mu l,
\hspace{0.3cm}
O_8=\frac{e^2}{16\pi^2}\sbar_L \gmu b_L \lbar \gamma^\mu l,
\hspace{0.3cm}
O_9=\frac{e^2}{16\pi^2}\sbar_L \gmu b_L \lbar \gamma^\mu \gfive l.
\label{eqn:hk3}
\end{eqnarray}
Here the term
$\ms\sbar_R i\sigma_{\mu\nu}(q^\nu/\qsq)b_L$
has been neglected compared to
$\mb\sbar_L i\sigma_{\mu\nu}(q^\nu/\qsq)b_R$.
The QCD corrected
Wilson coefficients $C_i(\mb)$ are dependent on the top quark mass.
In the standard model, the vector and axial vector current operators
$O_8$ and $O_9$ yield the dominant contributions to
the Hamiltonian (\ref{eqn:hk1})
and the contributions of the magnetic moment type operator
$O_7$ is less than 10 \% of $O_8$ and $O_9$.
The coefficient $C^{eff}_8(\mb)$ contains the contributions from
the $c\bar{c}$ continuum obtained from the electromagnetic penguin
diagram and the long distance contributions due to the $J/\psi$
and $\psi'$ poles
\cite{long1}, and therefore it is dependent on $\qsq$.
In general, the long distance contributions are significant.
However, they are quite small in the regions of $\qsq$
relevant for our analysis; $\qsq \gsimeq 0.6 \mbbar^2$.
(See e.g., figs 3(a), (b) of Lim {\it et. al} in Ref. \cite{long1}.)
The analytic expressions of the Wilson
coefficients and their numerical values
are given in Refs. \cite{grin}.
The hadronic matrix elements of the
magnetic moment type operator and the vector and
axial vector currents
are necessary to evaluate the Hamiltonian (\ref{eqn:hk1}).
The vector and axial vector currents are expressed in terms of the
form factors as,
\begin{eqnarray}
\hmkb{\gamma_{\mu}} &=& i g^{K^*} \epsilon_{\mu\nu\rho\sigma}
\epstar^{\nu}(\pppp)^{\rho}(\pmpp)^{\sigma},
\label{eqn:fk2}
\\
\hmkb{\gamma_{\mu}\gfive} &=& f^{K^*} \epstar_{\mu}
+\ap^{K^*}(\epp)(\pppp)_{\mu}
+\am^{K^*}(\epp)(\pmpp)_{\mu},
\label{eqn:fk1}
\end{eqnarray}
in the same way as in eqs. (\ref{eqn:f2}) and (\ref{eqn:f1}).
Again, only the term $f^{K^*} \epstar_i$
remains non-zero in the limit of zero recoil $\kstar$.
As for the magnetic moment type operator, since
$q^\nu = (\mbbar-\mkstar,\vec{{\bf 0}})+{\cal O}(\pks)$,
only the components
$\bar{s} i \sigma_{0i} b$ and $\bar{s}i \sigma_{0i}\gfive b$
are relevant in the same limit.
The hadronic matrix elements of these operators can be related
to those of the vector and axial vector currents (\ref{eqn:fk2}) and
(\ref{eqn:fk1}) by the static heavy quark approximation.
In this approximation, the $b$ quark in the $\bbar$ meson stays
on-shell throughout the reaction, and we can set
the equation of motion for $b$ quark, $\gamma_0 b=b$,
which leads to the relations \cite{iw}
\begin{eqnarray}
\hmkb{i \sigma_{0 i} } &=&  \hmkb{\gamma_{i}},
\label{eqn:hqa}
\\
\hmkb{i \sigma_{0i}\gfive } &=& - \hmkb{\gamma_{i}\gfive}.
\label{eqn:hqb}
\end{eqnarray}
The right-hand sides of eqs. (\ref{eqn:hqa}) and (\ref{eqn:hqb})
can be expressed in terms of the same form factors which appeared in
eqs (\ref{eqn:fk2}) and (\ref{eqn:fk1}),
and only the term $f^{K^*}\epstar_i$ remains non-zero in the zero recoil
limit.
Thus, the hadronic matrix elements required for $\bksll$ are
described only in terms of the form factor $f^{K^*}$ in the vicinity
$\pks \simeq 0$.
The $\qsq$-distribution of the decay width is given by
\begin{eqnarray}
\frac{d \Gamma(\bksll)}{d\qsq} &=&
|\vtb\vsts|^2 {\frac{\gf}{32\pi^3\mbbar^2}}|f^{K^*}|^2
(\frac{\alpha_{QED}}{4\pi})^2
2 (C^2_V + C^2_A)
\qsq \pks
+{\cal O}(\pks^3),
\label{eqn:qsq2}
\\
C_V &=& -C^{eff}_8(\mb)_{\qsq=\qsqmax} +
\mb \frac{(\mbbar-\mkstar)}{\qsqmax} C_7(\mb),
\hspace{0.5cm}
C_A = -C_9(\mb). \nonumber
\end{eqnarray}
Now we extract the $\vub$ from the $\qsq$-distributions
(\ref{eqn:qsq1}) and
(\ref{eqn:qsq2}).
Applying the $SU(3)$-flavor symmetry to the $\rho$
and $\kstar$ mesons. The question arises as to where we expect the
form factors $f^\rho$ and $f^\kstar$ to be nearly equal. This can be settled
experimentally by studying the $q^2$ distribution near
$\mbox{\boldmath $p$}_{\rho,\kstar}\to 0$
limit. For now, the best guess is that the
$SU(3)$-flavor
symmetry holds when $u$ and $s$ quarks in respective $b$ decays to be
at rest in the B meson rest frame,
\begin{eqnarray}
f^\rho(\qsqmr) =
f^{\kstar}(\qsqmk)
\label{eqn:su3}
\end{eqnarray}
which is expected to be valid in the region $\qsqmr=(\mbbar-\mrho)^2$ and
$\qsqmk=(\mbbar-\mkstar)^2$, respectively.
Then the ratio ${|\vub|^2}/{|\vtb\vsts|^2}$ is extracted as
\begin{eqnarray}
\frac{|\vub|^2}{|\vtb\vsts|^2}&=&\frac{\qsqmk}{\qsqmr}
(\frac{p_{\kstar}}{p_{\rho}})_{lim}
(\frac{\alpha_{QED}}{4\pi})^2
{2(C_V^2+C_A^2)}
\nonumber
\\
& &\cdot
[ \frac{d \Gamma(\brln)}{d \qsq}]_{\qsq \rightarrow \qsqmr}
/
[ \frac{d \Gamma(\bksll)}{d\qsq}]_{\qsq \rightarrow \qsqmk}.
\label{eqn:main}
\end{eqnarray}
Here
$({p_{\rho}}/{p_{\kstar}})_{lim}=
\sqrt{\mrho/\mkstar}$.
In the limit $\mbox{\boldmath $p$}_{\rho,\kstar} \rightarrow 0$, i.e.,
$\qsq \rightarrow \qsq_{max}$, the $\qsq$
distributions vanish due to the phase space suppression.
However, the numerical values of
the coefficients
of $\mbox{\boldmath $p$}_{\rho,\kstar}$ in eqs. (\ref{eqn:qsq1}) and
(\ref{eqn:qsq2})
can be precisely extracted in experiments.
In fact, CLEO collaboration has accurately
determined the value of $|V_{cb}|f(\qsqmax)$
for the process $\bbar \rightarrow \dstar l \bar{\nu} $ \cite{cleo1}.
In the similar manner, the right-hand side of eq. (\ref{eqn:main})
can be determined by experiments.
Expression (\ref{eqn:main}) is our main result.
In this expression, $\vsts$ itself is not directly measured,
however it is well determined by the unitarity condition, so
$|\vub|$ can be precisely evaluated.
Also once $|\vub|$ is known by some other methods,
$|\vsts|$ can be determined from eq. (\ref{eqn:main}).
Let us discuss the number of events needed for an analysis of this type.
{}From the figure shown in Lim $et.~al.$ \cite{long1}, we can read off
\begin{eqnarray}
&&\int^{.8}_{.6}ds\frac{dBR.(B\to e^+e^-+anything)}{ds}
\sim (3-5)\times 10^{-7},
\end{eqnarray}
where $s={q^2}/{m_b^2}$.
For a luminosity such that $3\times10^7~~B\bbar$ pairs are produced,
there should be $72-120$ $B\to e^+e^-~or~\mu^+\mu^-+anything$ events in the
kinematic region $.6<s<.8$ in one year of running the B factory.
In this kinematic region, $anything$ should be dominated by $K^*$.
If the higher end of the estimate is valid, barring unexpected background
or systematic problems, we hope to perform a
10\% level measurement of $|V_{ub}|$.
The theoretical uncertainty in the derivation of eq. (\ref{eqn:main})
lies in eq. (\ref{eqn:su3}), stemming from
the breaking of the $SU(3)$-flavor symmetry in the $\rho$ and
$\kstar$ mesons.
This is expected to be small.
For example, we may guess that
the ratio of the wave functions of
$\rho$ and $\kstar$ mesons is estimated by
the ratio of their decay constants\cite{hayakawa}:
\begin{eqnarray}
\frac{g_{\kstar K\pi}}{g_{\rho\pi\pi}}=1.08\pm .02.
\end{eqnarray}
The difference of the Fermi motion of the $b$ quark in the
decays $\brln$ and $\bksll$ may give rise to the error of
order $(\ms^2-m^2_u)/\{(M_{\bbar}-\mb)\mb\}$ \cite{bigi}.
An accurate computation of the ratio
$f^{\kstar}(\qsqmk)/f^\rho(\qsqmr) $ will allow a more precise
determination of $|\vub|$.
Corrections to the relations (\ref{eqn:hqa}) and
(\ref{eqn:hqb}) are of order $\Lambda_{QCD}/\mb$ in the
zero recoil $\kstar$ limit. The uncertainty due to these corrections
are significantly reduced in the level of the $\qsq$-distribution
(\ref{eqn:qsq2}), because
the contributions of the operator $O_7$
is numerically less than 10 \% of those of $O_8$ and $O_9$,
and accordingly we expect this error to be of order
$\Lambda_{QCD}/\mb\times 10\%\sim .4\%$ and is negligible.
In general, $C_{V,A}$ in eq. (\ref{eqn:main})
may be sensitive to the parameters of
new physics beyond the standard model.
This fact provides us with an interesting possibility that a value of
$|\vub|$ extracted in our strategy will play a role
in probing for new physics, by comparing values of $|\vub|$ determined
by other methods.
We have used the $\qsq$-distributions of $\brln$ and
$\bksll$ to determine $|\vub|$.
Studies on the
forward-backward asymmetry of the leptons \cite{moro}
and the polarization of $\rho$ and $\kstar$ mesons may be also useful.
The forward-backward asymmetry is described by the form factor
$f^\rho g^\rho$
in the decay $\brln$, and is described by $(f^{\kstar})^2$,
$(g^{\kstar})^2$ and $f^\kstar g^\kstar$ for
$\bksll$.
The terms $(f^{\kstar})^2$ and $(g^{\kstar})^2$
come from the magnetic moment type operator
$O_7$, when its hadronic matrix elements are
related to those of $O_8$ and $O_9$ in the static
heavy quark approximation.
Along the similar line, an analysis similar to ours
can be made using the
radiative decay $\bksg$ \cite{bsg1}.
However the theoretical prediction of this decay rate
suffers from the uncertainty due to the large recoil momentum of the
$\kstar$ meson and the long distance contributions \cite{pa}.
One of us (A.~Y) would like to thank K-I.~Izawa, N.~Kitazawa,
T.~Morozumi, M.~Tanabashi and S.~Uno for useful discussions, and
L.~T.~Handokoo for sending his computer program of the QCD corrected
Wilson coefficients.


\begin{thebibliography}{99}
\bibitem{km}
M.~Kobayashi and K.~Maskawa, \prog{49}{1973}{652}.
\bibitem{sanda1}
A.~B.~Carter and A.~I.~Sanda, \prl{45}{1980}{952};
\prd{23}{1981}{1567};
I.~I.~Bigi and A.~I.~Sanda, \prd{29}{1984}{1393};
\npb{193}{1981}{85}; Comments Nucl. Part. Phys. 14, 149 (1985);
\npb{281}{1987}{41};
I.~Dunietz and J.~Rosner, \prd{34}{1986}{1404};
D.~Du, I.~Dunietz and D.Wu, \prd{34}{1986}{3414}.
\bibitem{vubc}
There investigated several methods for the determinations of $|\vub|$.
Inclusive processes are studied in,
P.~Ball, V.~M.~Braun and H.~G.~Dosch,
\prd{48}{1993}{2110};
C.~S.~Kim, P.~Ko, D.~Hwang and W.~Namgung,
\prd{50}{1994}{5762};
B.~Blok and T.~Mannel, \prd{51}{1995}{2208};
C.~S.~Kim and A.~D.~Martin, \plb{345}{1995}{296}.
Semileptonic decays are used in,
G.~Kramer, T.~Mannel and G.~A.~Shuler,
\zeitc{51}{1991}{649};
D.~Du and C.~Liu, \prd{50}{1994}{4558};
H.~Li and H.~L.~Yu, \prl{72}{1995}{4388};
N.~Kitazawa, \plb{349}{1995}{541}. This analysis is
based on the effective Lagrangian
obtained in N.~Kitazawa and T.~Kurimoto, \plb{323}{1994}{65};
A.~Datta, UH-511-825-95, hep-ph-9504429;
M.~Oda, M.~Ishida and S.~Ishida, NUP-A-94-7.
Non-leptonic decays are analyzed in,
I.~Dunietz and J.~Rosner, CERN-TH-5899-90;
D.~Choudhury, D.~Indumati, A.~Soni and S.~U.~Sankar,
\prd{45}{1992}{217};
N.~G.~Deshpande and C.~O.~Dib, \plb{319}{1993}{313}.
The use of the radiative $B$ decays has not been seriously
investigated in these analyses.
\bibitem{inamilim}
T.~Inami and C.~S.~Lim, \prog{65}{1981}{297}.
\bibitem{long1}
C.~S.~Lim, T.~Morozumi and A.~I.~Sanda, \plb{218}{1989}{343};
N.~G.~ Deshpande, J.~Trampetic and K.~Panose, \prd{39}{1989}{1461};
P.~J.~O'Donnell and H.~K.~K.~Tung, \prd{43}{1991}{R2067}.
\bibitem{grin}
B.~Grinstein, M.~J.~Savage and M.~B.~Wise, \npb{319}{1989}{271};
M.~Misiak, \npb{393}{1993}{23};
M.~Ciuchini, E.~Franco, G.~Martinelli and L.~Reina,
\npb{415}{1994}{403}. Misiak denotes our $O_8$ and $O_9$ as
${\tilde O}_9$ and ${\tilde O}_{10}$, respectively.
Strategies to determine the Wilson coefficients from experiments
of rare $B$ decays are discussed
e.g., in C.~Greub, A.~Ioannissian and D.~Wyler, \plb{346}{1995}{149};
A.~Ali, G.~F.~Guidice and T.~Mannel,
CERN-TH7346, hep-ph-9408213.
\bibitem{iw}
N.~Isgur and M.~B.~Wise, \prd{42}{1990}{2388}.
\bibitem{cleo1}
CLEO collaboration, \prd{47}{1993}{791};
\prd{51}{1994}{1014}. Their form factor $A_1(\qsq)$ is equal to
our $(\mbbar+m_{\dstar}) f^{\dstar}(\qsq)$.
\bibitem{hayakawa}
Review of Particle Properties, \prd{50}{1994}{1173}; See also
M.~Hayakawa, T.~Kurimoto and A.~I.~Sanda, \prog{92}{1994}{377}.
It is interesting that the similar estimate is obtained by
Burdman and Donoghue in a different context;
G.~Burdman and J.~F.~Donoghue, \plb{270}{1991}{55}.
Using the Bauer-Stech-Wirbel model,
they estimate that the ratio $f^{\kstar}(0)/f^\rho(0)$
is 1.04$\times$1.10.
\bibitem{bigi}
I.~I.~Bigi, M.~A.~Shifman, N.~G.~Uraltsev and
A.~L.~Vainstein, \ijmp{9}{1994}{2467}.
\bibitem{moro}
A.~Ali,~T. Mannel and T.~Morozumi study
the forward-backward asymmetry in the inclusive level
$b \rightarrow s l\lbar$;
A.~Ali, T.~Mannel and T.~Morozumi, \plb{273}{1991}{505}.
\bibitem{bsg1}
Burdman and Donoghue relate the decays $\bksg$ and $\brln$
at a particular point in the Dalitz plot;
Burdman and Donoghue, in Ref. \cite{hayakawa}.
See also
P.~J.~O'Donnell and H.~K.~K.~Tung, \prd{48}{1993}{2145}.
\bibitem{pa}
E.~Golowich and S.~Pakvasa, \plb{205}{1988}{393};
N.~G.~Deshpande, J.~Trampetic and K.~Panose, \plb{214}{1988}{467}.
\end{thebibliography}
\end{document}